\documentclass{svjour3}                     
\smartqed  
\usepackage{graphicx}
\usepackage{braket}

\usepackage{amssymb}
\usepackage{tensor}

\newcommand{\ve}[1]{\mathbf{#1}}

\begin{document}

\title{Semiclassical approximation of the Wheeler-DeWitt equation:
  arbitrary orders and the question of unitarity}

\titlerunning{Semiclassical approximation of the Wheeler-DeWitt
  equation}        

\author{Claus Kiefer        \and David Wichmann
}

\institute{C. Kiefer \at
              Institute for Theoretical Physics,
              University of Cologne\\
              Z\"ulpicher Stra\ss e 77,
              50937 K\"oln, Germany\\
              \email{kiefer@thp.uni-koeln.de}  
           \and
D. Wichmann \at
Institute for Theoretical Physics,
              University of Cologne\\
              Z\"ulpicher Stra\ss e 77,
              50937 K\"oln, Germany
 \\ \email{d.wichmann@uu.nl} \\ 
  \emph{Present address:} \\
  Institute for Marine and Atmospheric Research, Utrecht University \\
  Princetonplein 5, 3584 CC Utrecht, Netherlands 
}

\date{Received: date / Accepted: date}

\maketitle

\begin{abstract}
We extend the Born-Oppenheimer type of approximation scheme for the
Wheeler-DeWitt equation of canonical quantum gravity to arbitrary
orders in the inverse Planck mass squared. 
We discuss in detail the origin of unitarity violation in this scheme
and show that unitarity can be restored by an appropriate modification
which requires back reaction from matter onto the gravitational
sector. 
In our analysis, we heavily rely on the
gauge aspects of the standard Born-Oppenheimer scheme in molecular
physics.

\keywords{Canonical Quantum Gravity \and Semiclassical Approximation \and
  Wheeler-DeWitt Equation \and Born-Oppenheimer Approximation} 

\end{abstract}

\section{Introduction}
\label{intro}

In the search for a more fundamental theory, it is of the utmost
importance to understand the connection of the new theory with 
existing and empirically established theories. This holds, in
particular, for the goal of constructing a theory of quantum
gravity. By now, various approaches exist, but there is no agreement on
which is the right one \cite{oup}. One necessary requirement for any
approach is that its semiclassical limit contains classical gravity
and quantum field theory in a background spacetime. Understanding this 
limit could also enable one to go beyond it and calculate quantum
gravitational correction terms that can potentially be observed and
could thus serve as a test for the theory.

One conservative but promising approach is canonical quantum gravity
in the metric variables. Unlike, for example, string theory, this is
not a unified theory of all interactions. But it is still expected
to give reliable information about gravity in the quantum realm
\cite{grg}: if one rewrites Einstein's equations into Hamilton-Jacobi
form and formulates (in the spirit of what Schr\"odinger did for
mechanics in 1926) quantum wave equations from which the
Hamilton-Jacobi form can be recovered in the WKB limit, one
necessarily arrives at the equations of quantum
geometrodynamics. These equations should thus hold, at least
approximately, as long as the linear structure of quantum theory
remains valid. It should then be possible to extract meaningful
predictions from this framework such as the ones discussed here,
which concern quantum gravitational {\em corrections} to the limit of
quantum field theory in curved spacetime.  

The central equations of canonical quantum gravity are four local
constraints -- the 
Wheeler-DeWitt equation and three momentum (diffeomorphism)
constraints. Their semiclassical limit has already been studied in a
variety of ways, see \cite{oup} and the references therein. What we
want to add here are essentially two things. First, we want to extend
the previous expansion scheme of \cite{KS91} to arbitrary orders in the
appropriate parameters. And second, we want to comment on the issue
whether quantum gravitational correction terms break the usual
unitarity of quantum theory or not. The latter point is clarified by
drawing an analogy with the gauge structure of the Born-Oppenheimer
approach in molecular physics. 

To be more concrete, we consider the
Wheeler-DeWitt equation in the following form:\footnote{There is no
  need to address the momentum constraints here, because their sole
  purpose is to guarantee the invariance of the wave functional under
  three-dimensional coordinate transformations.} 
\begin{equation}
 \left[ -\frac{\hbar^2}{2M}\left(G_{ab}\frac{\delta^2}{\delta h_a \delta
       h_b} + g_a \frac{\delta}{\delta h_a}\right)+ MV(h_a) +
   H_{\rm m}(h_a,\phi)\right]\Psi[h_a,\phi] = 0 \label{eq:WDW-Equation}. 
\end{equation}
Here, indices $a, b, \ldots$ represent a symmetric double index
and $h_a$ denotes the spatial three-metric; $G_{ab}$ is the DeWitt
metric. The variable
$\phi$ represents a (bosonic) matter
field with Hamiltonian $H_{\rm m}$ that only depends parametrically on
$h_a$. The parameter $M:={c^2}/{32\pi G}\approx 1.34\times
10^{25}\ {\rm kg}/{\rm m}$ is related to the square of the
(reduced) Planck mass $M_{\rm P}=\sqrt{\hbar c/8\pi G}$
by $M=cM_{\rm P}^2/4\hbar$; $M$ resp. $M/\hbar\approx 1.27\times
10^{59}\ {\rm s}/{\rm m}^3$
will be the appropriate formal parameter for
the Born-Oppenheimer scheme below. Finally, $V:= -2c^2 \sqrt{\det h_a}R$ denotes the gravitational
potential with $R$ as the three-dimensional Ricci scalar.
The functions $g_a$ are introduced to parameterize factor ordering
ambiguities. In the arguments of the wave functionals, we will
often suppress the indices of the three-metric for simplicity.

The limit of quantum field theory in curved spacetime has been derived
by two different but closely related expansion schemes. One is a direct
expansion with respect to the parameter $M$ in (\ref{eq:WDW-Equation})
\cite{KS91}, the other is a more or less direct application of the
molecular Born-Oppenheimer scheme \cite{BFV96,BK98}; its main
difference lies in the treatment of back reaction on the gravitational
sector and the preservation or violation of unitarity in the matter sector.
 A recent comparison can be found in \cite{KTV17} and in the Appendix
 of \cite{Bini13}. 

In the first approach,
the correct limits of classical gravity and the functional
Schr\"odinger equation of quantum field theory in a fixed curved
background can be obtained from (\ref{eq:WDW-Equation}) by
making the ansatz \cite{KS91,oup}
\begin{equation}
 \Psi[h,\phi] = \chi[h] \psi [h,\phi] \label{eq:BO_Ansatz},
\end{equation}
together with a WKB-like expansion in $M^{-1}$ for $\chi$ and
$\psi$.\footnote{Note that Born and Oppenheimer in their classic paper
  \cite{BornOppenheimer27} did not perform a WKB-like expansion, but a
  Taylor-series-like expansion in the small parameter
  $\kappa=\sqrt[4]{\frac{m_{\rm electron}}{m_{\rm nucleus}}}$.}
For the Wheeler-DeWitt equation, this approach resembles the
traditional Born-Oppenheimer ansatz of molecular physics with
 zero total energy.\footnote{An exact description would require in
   addition a sum over a complete set of eigenstates $\psi_n$, but we
   stay in the regime of the \emph{adiabatic} approximation where the
   off-diagonal terms of the Mead-Berry connection are neglected,
   similar to the ansatz of Born and Oppenheimer
   \cite{BornOppenheimer27}. This neglection can be justified by the
   process of decoherence \cite{deco}.} There is, however, an important
 difference to molecular physics. We use the ansatz in
 (\ref{eq:BO_Ansatz}) to derive a semiclassical limit for
 $\psi$ alone, accompanied by the recovery of a semiclassical (WKB)
 time parameter (more precisely, a local many-fingered time) through a
 corresponding functional $\chi$.
We also want to interpret $\psi$ by itself as a meaningful
 wave functional. We do so by deriving its functional Schr\"odinger
 equation from (\ref{eq:WDW-Equation}), which results from a
 choice of functional $\chi$. We then face the problem of choosing a
 reasonable $\chi$. 

This problem becomes more apparent by noting that the solution of
(\ref{eq:WDW-Equation}), $\Psi$, is invariant under a rescaling
of $\chi$ and $\psi$ of the form $\chi \rightarrow e^{A} \chi, \; \psi
\rightarrow e^{-A}\psi$ for an arbitrary complex valued functional
$A[h]$. Since in the semiclassical approximation to quantum gravity a
time parameter is defined through a 
functional depending on $h$ (see below), the freedom to choose a
``gauge'' $A[h]$ will 
influence the time evolution of both $\chi$ and $\psi$ and can thus have 
consequences for unitarity. By unitarity we here mean the conservation
of the standard Schr\"odinger inner product for the matter wave function $\psi$ with
respect to semiclassical (WKB) time.
Note that the gauge freedom of $A[h]$ is the same as the one
that leads to the gauge theory of molecular physics (see, for example,
\cite{Bohm03} and \cite{Baer06}), although there one restricts to
transformations that leave $\psi$ normalized to unity, that is, $A$
is purely imaginary, and the gauge group is the unitary group, see
section~2 for more details. As we want
to interpret such a $\psi$ and the equation governing its evolution
physically, we need a guiding principle for the choice of an
appropriate gauge $A[h]$. Such a principle will be proposed and applied
in the present paper.

Our paper is organized as follows.
In section~2, we outline the underlying gauge structure of the
Born-Oppenheimer scheme, which is crucial for our discussion. 
In section \ref{sec:traditional}, we follow the ansatz of
\cite{KS91} where $\chi$ is taken to be a solution to the
vacuum Wheeler-DeWitt equation. Within this framework, we derive a formal
expression containing {\em all} quantum gravitational corrections at
successive orders 
of $M^{-1}$. In section \ref{sec:UnitaryPreservation}, we then show how
to choose $\chi$ such that $\psi$ obeys a unitary time evolution,
and demonstrate the similarity of the calculations in the functional
Schr\"odinger picture for $\psi$ to the traditional Born-Oppenheimer
approach, where one considers the equation for $\chi$ (the ``nuclear
wave function'') after restricting the gauge group to the unitary group
(i.e. with $A[h]$ purely imaginary). Both of these approaches will yield
the correct limits of classical gravity and quantum field theory in a
fixed curved background spacetime, but the quantum gravitational
corrections will turn out to be different.
The last section contains a brief summary and an outlook on possible
applications.

\section{Lessons from molecular physics}

In this section, we start by following the standard treatment
presented, for example, in \cite{Bohm03} and \cite{Baer06} and
continue by making further elaborations which are relevant for the
unitarity issue. 

In the Born-Oppenheimer approximation of molecular physics, one
considers molecules which consist of interacting nuclei and
electrons. Because the nuclei are more massive and move slower than
the electrons, one can divide the total system into a slow part
(nuclei) and a fast part (electrons). These two parts are only weakly
coupled and suitable for a perturbative treatment with a naturally
arising small parameter defined by the mass ratio of electrons and
nuclei.\footnote{While the traditional Born-Oppenheimer approximation
  is based on this separation, there exist approaches that treat nuclei
  and electrons on the same footing \cite{Matyus}. It would be
  interesting to apply them to quantum gravity.}

The total Schr\"odinger equation for a molecule reads
\begin{equation}
 {\rm i} \hbar \partial_t \ket{\Psi} = \left(T_{\rm nuc} + T_{\rm el} + V
 \right)\ket{\Psi}  \label{eq:HamiltonianMolecule}. 
\end{equation}
Here, $T_{\rm nuc}$ and $T_{\rm el}$ denote the kinetic energy operators of
nuclei and electrons, respectively, and $V$ stands for all nuclei-nuclei,
electron-electron, and nuclei-electron interactions.

For every fixed nuclear position $R$, we can choose a complete set
$\ket{n(R)},\; n=1, \dots, N,$\footnote{In realistic cases, $N$ will be
  infinite.} for the electronic part of the quantum state. In these
states, $R$ appears as a parameter only. To be more general, we should
also include time $t$ into this state. We hence have a basis
$\ket{n(R,t)}$ for each fixed nuclear configuration $R$ and fixed time
$t$. Let us denote the nuclear configuration space as $\mathcal{S}$,
such that the whole `background space' is $\mathcal{S}\otimes
\mathbb{R}$. 

By the assumption of completeness for every configuration $R$ we can
write for the total state
\begin{equation}
 \ket{\Psi(t)} = \sum_n \int dR' \chi^n(R',t) \ket{R'}
 \ket{n(R',t)} \label{eq:Expansion} 
\end{equation}
with a set of components $\chi^n(R,t)$. 
Defining
\begin{equation}
\label{ket-psi}
 \ket{\psi(R,t)}:= \braket{R|\Psi} = \sum_n  \chi^n(R,t)\ket{n(R,t)}, \nonumber
\end{equation}
we get by multiplying
(\ref{eq:HamiltonianMolecule}) with $\bra{R}$ and using 
(\ref{eq:Expansion}) the following Schr\"odinger equation for 
$\ket{\psi(R,t)}$:
\begin{equation}
 {\rm i} \hbar \partial_t \ket{\psi(R,t)}  = \left(
   \frac{-\hbar^2}{2M}\nabla^2_R + H_{0}(R,t)\right)
 \ket{\psi(R,t)} \label{eq:TimeDepSchroedinger}, 
\end{equation}
where
\begin{equation}
\label{H0Rt}
 H_{0}(R,t) := T_{\rm el}(R) + V(R). \nonumber
\end{equation}
The mass $M$ denotes some average mass (e.g. the reduced mass) of the
nuclei, and the label $R$ a vector of all nuclear coordinates, which
are in general mass weighted. For simplicity, we abbreviate the state
$\ket{n(R,t)}$ by 
$\ket{n}$. If we define
\begin{eqnarray}
\label{column}
 \chi &=& (\chi^1, \chi^2, \dots)^T, \nonumber \\
 \ve{e}&=& (\ket{1},\ket{2},\dots) ,
\end{eqnarray}
we can write $\ket{\psi}$, Eq. (\ref{ket-psi}), as 
\begin{equation}
 \ket{\psi} = \ve{e}\chi \nonumber.
\end{equation}
Like any such product, $\ket{\psi}$ possesses a
${\rm GL} (N,\mathbb{C})$ invariance of the form
\begin{eqnarray}
 \chi &\rightarrow& C^{-1} \chi \nonumber \\
 \ve{e} &\rightarrow& \ve{e}\ C
\end{eqnarray}
for an arbitrary non-degenerate matrix $C(R,t) \in {\rm GL} (N,\mathbb{C})$.\\

The gauge group of molecular physics is
obtained by choosing $\{\ket{n}\}$ to form a local
\emph{orthonormal} basis; in this way, ${\rm GL} (N,\mathbb{C})$
is reduced to the unitary group ${\rm U}(N,\mathbb{C})$. The usual
Born-Oppenheimer approach uses for 
the $\{\ket{n}\}$ the \emph{stationary}\footnote{These are the states without the
  time-dependent phase $e^{-\frac{\rm i}{\hbar}E(R)t}$.} orthonormalized
eigenstates of the `electronic' Hamiltonian $H_0$ defined in (\ref{H0Rt}). This
corresponds to choosing a particular representative of the unitarily
equivalent set of orthonormal bases $\{\ket{n}\}$.

Let us consider the time-dependent Schr\"odinger equation
(\ref{eq:TimeDepSchroedinger}). Inserting $$\ket{\psi} = \sum_m \chi^m
\ket{m}$$ (which is a shorthand writing of (\ref{ket-psi})) and
contracting with $\bra{n}$ from the left, we get by the 
orthonormality of the $\ket{n}$ the following equations for the $\chi^n$: 
 \begin{equation}
  {\rm i} \hbar \dot{\chi}^n + i \hbar \sum_m \chi^m\braket{n|\partial_t|m}
  = \sum_m \left(-\frac{\hbar^2}{2M}\braket{n|\nabla^2_R|m} +
    \tensor{H}{_0 ^n _m} \right)\chi^m \label{eq:HamiltonianNuclei}, 
 \end{equation} 
 where the Laplacian $\nabla^2_R$ acts on everything on its right, and
 $\tensor{H}{_0 ^n _m}:= \braket{n|H_{0}|m}$. It is appropriate to
 define the following quantities:
 \begin{eqnarray}
  \tensor{\tau}{^n _i _m} &:=&\braket{n|\partial_i |m} \nonumber \\
  \tensor{\epsilon}{^n_m} &:=&\braket{n|\partial_t + \frac{\rm i}{\hbar}
                               H_0| m} \label{eq:ConnectionDefs}. 
 \end{eqnarray}  
 The $\tau$ are the usual Mead-Berry 
connections \cite{Bohm03,Baer06}.\footnote{Sometimes
  the imaginary unit is included in the definition.} As we will show
here, see (\ref{eq:SpacetimeConnection}) below, $\tau$ and 
 $\epsilon$ together lead to a single connection on $\mathcal{S}\otimes
 \mathbb{R}$. We note that both $\tau$ and $\epsilon$ are skew-hermitian; 
considered as a connection, they thus have the unitary group as gauge
group.

If we use again the column vector notation introduced in
(\ref{column}), 
we find for $\chi$ the equation
\begin{equation}
 {\rm i} \hbar (\partial_t + \epsilon)\chi = \left( \frac{-\hbar^2}{2M}
   (\nabla + \tau)^2 \right) \chi \label{eq:GaugeMolecule}. 
\end{equation}
Note that $\tau$ and $\epsilon$ are now matrices that appear as 
connections on $\mathcal{S}\otimes \mathbb{R}$.\footnote{To our knowledge,
  one has in the literature so far only interpreted 
  $\tau$ as a connection.} Equation (\ref{eq:GaugeMolecule}) is a
consistency condition on the nuclear wave functions $\chi$, which has
to be satisfied if the 
$\ket{n}$ are orthonormal for every point in  $\mathcal{S}\otimes
\mathbb{R}$.

We note that in the usual Born-Oppenheimer framework, one
chooses \emph{time-independent} eigenstates with
$H_0\ket{n}=e_n\ket{n}$ (without the factor $e^{-{\rm i}/\hbar e_n(R)
  t}$). For these states $\partial_t \ket{n}=0$, and hence
$\tensor{\epsilon}{^n_m} = \frac{\rm i}{\hbar}\delta^n_m e_n$. Equation
(\ref{eq:GaugeMolecule}) then reduces to the time-dependent form of the
Born-Oppenheimer approach: 
\begin{equation}
 {\rm i} \hbar \partial_t \chi = \left( \frac{-\hbar^2}{2M} (\nabla +
   \tau)^2 + e \right) \chi, \label{eq:TraditionBOExpr}  
\end{equation}
with $e = {\rm diag}(e_1, e_2, \dots)$. One usually also considers the
stationary Schr\"odinger equation for the full state. In this case,
the full eigenvalue $E$ occurs on the left-hand side of 
(\ref{eq:HamiltonianNuclei}). Equation (\ref{eq:GaugeMolecule})
then reduces to the well-known form \cite{Bohm03,Baer06}
\begin{equation}
 E \chi = \left( \frac{-\hbar^2}{2M} (\nabla + \tau)^2 + e \right)
 \chi \label{eq:GaugeMoleculeStationary}. 
\end{equation}
One can interpret (\ref{eq:TraditionBOExpr})  and
(\ref{eq:GaugeMoleculeStationary}) as resulting from a choice of gauge in
(\ref{eq:GaugeMolecule}). All these equations are unitarily
related. 

Let us check that the connections defined in (\ref{eq:ConnectionDefs})
indeed transform correctly under a unitary transformation (as also
done e.g. in \cite{Bohm03} for $\tau$ alone). Writing again
the basis $\ket{n}$ as a row vector, and $\ve{e}=(\ket{1},
\ket{2},\dots)$, we have again $\ket{\psi} =
\ve{e}\chi$. The modified Schr\"odinger equation
(\ref{eq:GaugeMolecule}) should be invariant under all transformations
that leave $\ve{e}$ orthonormal (in the quantum sense), that is,
under unitary transformations depending on nuclear coordinates and time. For
a unitary transformation of the form 
\begin{equation}
 \ket{k} \rightarrow \ket{k} \tensor{U}{^k _n} \nonumber,
\end{equation}
the corresponding bra transforms such that the orthonormality
$\braket{n|m}=\delta^n_m$ is preserved: 
\begin{equation}
 \bra{k}  \rightarrow \tensor{U}{^*^k _n}\bra{k} = \tensor{U}{^\dag ^n
   _k}\bra{k} \nonumber. 
\end{equation}
In order for $\ket{\psi}$ to be invariant
under such a transformation, we have to demand that $\chi$ transforms
just as $\bra{n}$: 
\begin{equation}
  \chi^k \rightarrow \tensor{U}{^\dag ^n _k}\chi^k \nonumber.
\end{equation}
In vector notation this reads
\begin{eqnarray}
 \ve{e} \rightarrow \ve{e} U, \nonumber \\
 \chi \rightarrow U^\dag \chi \nonumber.
\end{eqnarray}
Let us now see how the $\tau$ and $\epsilon$-matrices
transform under gauge transformations. We find
\begin{eqnarray}
 \tensor{\tau}{^n_m} &\rightarrow &\tensor{\tilde{\tau}}{^n_m} =
                                    \tensor{U}{^\dag
                                    ^n_p}\braket{p(R)|\nabla
                                    |q(R)}\tensor{U}{^q _m} \nonumber
  \\ 
 &=& \tensor{U}{^\dag ^n _p}\tensor{\tau}{^p _q} \tensor{U}{^q _m} +
     \tensor{U}{^\dag^m_p} \nabla \tensor{U}{^p_m}, \nonumber \\ 
 \tensor{\epsilon}{^n_m} &\rightarrow& \tensor{\tilde{\epsilon}}{^n_m}
                                       = \tensor{\epsilon}{^n_m} +
                                       \tensor{U}{^\dag^n_k} \partial_t
                                       \tensor{U}{^k_n} \label{eq:GaugeTrafo}, 
\end{eqnarray}
which are the correct transformation laws for a connection. If we now
define a `spacetime connection' $\omega$ as 
\begin{eqnarray}
 \tensor{\omega}{^m _0 _n} &=& \tensor{\epsilon}{^m_n} \nonumber \\
 \tensor{\omega}{^m _i _n} &=& \tensor{\tau}{^m _i
                               _n}, \label{eq:SpacetimeConnection} 
\end{eqnarray}
we see that $\omega$ transforms as a connection on
$\mathcal{S}\otimes \mathbb{R}$. We note again that
the gauge group is the unitary group, as the connection $\omega$ is
skew-hermitian. This is the same group that leaves the form of
equation (\ref{eq:GaugeMolecule}) invariant.  

A natural question to ask is if the fibre bundle defined by the
connection in (\ref{eq:SpacetimeConnection}) is trivial,
that is, if the theory defined by (\ref{eq:GaugeMolecule}) is
unitarily equivalent to a theory with $\tau = \epsilon =0$. 
It turns out that the answer is {\em no}. 
If we consider the curvature  $\Theta$ associated with the  spacetime
connection (\ref{eq:SpacetimeConnection}), we find after some calculations
that its non-zero components read $$\tensor{\Theta}{^m _n _i _0}
= \frac{\rm i}{\hbar} D_i \tensor{H}{_0 ^m_n},$$ 
where $D_i$ denotes the covariant derivative with respect to the
connection defined by $\omega$. Even if we choose a frame with
$\tau=0$ (and thus have $D_i = \partial_i$), we see that $\Theta$
would only vanish if $H_0$ was 
independent of $R$, which is clearly not the case.

How do we interpret the result $\Theta\ne0$? It means that there is no choice
of frame, that is, of orthonormal bases $\ket{n}$, such that $\chi$ obeys
the equation of free nuclei. This is physically very natural: an
interacting theory is not unitarily equivalent to a free theory. With
a view on quantum gravity, one should note, however, that we actually
started with a ${\rm GL}(N,\mathbb{C})$ invariance, so using this
larger invariance one can certainly choose $\chi$ 
to obey a free Schr\"odinger equation. Yet
this would spoil the normalization of the $\ket{n}$. In other words:
 if we put $\chi$ in a state with $\tau = \epsilon = 0$,
normalization of the $\ket{n}$ cannot be imposed for all configurations on
$\mathcal{S}\otimes \mathbb{R}$. This will become important for the
situation in quantum gravity, to which we will now turn.

\section{Semiclassical limit with vacuum gravity background} 
\label{sec:traditional}

\subsection{Quantum gravitational corrections to the functional
  Schr\"odinger equation} 
\label{subsec:VacuumAllOrders}

In this subsection, we will extend the scheme developed in \cite{KS91}
to higher orders. We start by defining the quantity
\begin{equation}
 \rho_\chi := \frac{1}{\chi}\left[
   -\frac{\hbar^2}{2M}\left(G_{ab}\frac{\delta^2}{\delta h_a \delta h_b} +
     g_a \frac{\delta}{\delta h_a}\right)+ MV \right]
 \chi, \label{eq:DefinitionRho} 
\end{equation}
which corresponds to the pure gravitational part in (\ref{eq:WDW-Equation}).
The scheme employed in \cite{KS91} corresponds to choosing
a wave functional $\chi$ (i.e. fixing a gauge) in (\ref{eq:BO_Ansatz}) with
$\rho_\chi=0$. That is, the arising picture is one in which the
semiclassical background is a solution of the vacuum Einstein field
equations on which the quantum matter fields described by $\psi$
propagate. One can, of course, generalize the scheme in order to
accommodate matter degrees of freedom that contribute to $\rho_\chi=0$
and thus represent a semiclassical matter part. In such a case, the
approximation scheme does not proceed with respect to $M^{-1}$, but
with respect to another appropriate parameter; recent examples are
the cases of de~Sitter inflation \cite{BKK16a} and of slow-roll
inflation \cite{BKK16b}. 

It is important to note that even though $\rho_\chi=0$
resembles the vacuum Wheeler-DeWitt equation, one has to be careful in
interpreting this $\chi$ as a physically meaningful functional within
our framework. For example, it would be hard to interpret a  similar
choice in molecular physics: without the interactions of electrons and nuclei,
no stable molecule could exist. For the gravitational case, the
situation is not as severe: it is certainly consistent to have a
vacuum solution of the Einstein equations on which the
(metric-dependent) matter fields evolve. Still, the choice
$\rho_\chi=0$ neglects possible back reaction terms from the matter
sector to the `background' part.   

 If we make a WKB-like ansatz analogously to
\cite{KS91},
\begin{equation}
 \chi = \exp\left[\frac{{\rm i}M}{\hbar} \sum_{j=0}^\infty
   \left(\frac{\hbar}{{\rm i} M}\right)^j 
   \sigma_j\right] =: \exp \left[\frac{{\rm i}M}{\hbar}(\sigma_0 +
   P)\right] \label{eq:WKBChi}, 
\end{equation}
that is,
\begin{equation}
P:=\sum_{j=1}^{\infty}\left(\frac{\hbar}{{\rm i}M}\right)^j\sigma_j,
\end{equation}
we get at orders $M^1$ and $M^0$ of the equation $\rho_\chi=0$ the
following equations, respectively,
\begin{eqnarray}
 \frac{1}{2}G_{ab}\frac{\delta \sigma_0}{\delta h_a}\frac{\delta
  \sigma_0}{\delta h_b} + V &=&0, \label{eq:HamiltonJacobi} \\ 
 G_{ab}\frac{\delta \sigma_0}{\delta h_a}\frac{\delta \sigma_1}{\delta
  h_b} +
  \frac{1}{2}G_{ab}\frac{\delta^2 \sigma_0}{\delta h_a \delta h_b}
+\frac{1}{2} g_a \frac{\delta \sigma_0}{\delta h_a}
                            &=&0  \label{eq:VanVleck}.  
\end{eqnarray}  
Equation (\ref{eq:HamiltonJacobi}) is the Hamilton-Jacobi equation for
vacuum gravity and is equivalent (because it is a functional equation)
 to all Einstein field equations
\cite{Gerlach69}. Note, however, that this equation only determines
the magnitude of the gradient of $\sigma_0$ (in terms of the DeWitt
metric), not its direction. 

One can now define a local time derivative
similar to \cite{KS91} as\footnote{This $\tau$ should not be confused
  with the Mead-Berry connection of the last section.} 
\begin{equation}
 \frac{\delta}{\delta \tau} := G_{ab}\frac{\delta \sigma_0}{\delta
   h_a}\frac{\delta}{\delta h_b} \label{eq:DefTimeDerivative}. 
\end{equation}
The many-fingered time $\tau(x)$ is not a scalar, but becomes one
after spatial integration \cite{oup}.
With this definition, it follows from 
(\ref{eq:HamiltonJacobi}) that the $\tau$-$\tau$ component of the DeWitt
metric becomes $G_{\tau \tau} = -{1}/{2V}$.\footnote{It should be
  noted that with this definition of the time derivative, 
  (\ref{eq:HamiltonJacobi}) leads to $\sigma_0 = - 2 \int
  V {\rm d} \tau$, which differs from the usual WKB case by having
  $V$ instead of its square root.} It is interesting
to note that all orders of $\rho_\chi=0$ will lead to
equations that determine only the gradients of the other $\sigma_j$'s
along the $\tau$-direction, $\delta \sigma_j/\delta \tau$, and
not the other components; this can be seen from (\ref{eq:VanVleck})
and from Table~1. The $\sigma_j$ for $j>1$ are hence only defined up to the
addition of an arbitrary $\tau$-independent functional. Such
additional terms do not follow from the previous orders, so their form
depends solely on the boundary conditions \cite{BK98}.

{\large
\begin{table}[h!]
\begin{tabular}{|c|c|}
\hline
\textbf{Order} & \textbf{Equation for $\sigma_j$}\\ 
\hline
$M$ & 
                                                      $\frac{1}{2}G_{ab}
                                                      \frac{\delta
                                                      \sigma_0}{\delta
                                                      h_a}
                                                      \frac{\delta
                                                      \sigma_0}{\delta
                                                      h_b} + V = 0$ \\   
\hline 
$M^0$ & 
                                       $G_{ab} \frac{\delta
                                         \sigma_0}{\delta h_a}
                                         \frac{\delta \sigma_1}{\delta
                                         h_a} + \frac{1}{2}G_{ab}
                                         \frac{\delta^2
                                         \sigma_0}{\delta h_a \delta
                                         h_b} +
                                         \frac{1}{2}g_a\frac{\delta
                                         \sigma_0}{\delta h_a}= 0$ \\ 
\hline
$M^{-1}$ & $G_{ab}
                                                     \frac{\delta
                                                     \sigma_0}{\delta
                                                     h_a}\frac{\delta
                                                     \sigma_2}{\delta
                                                     h_b} +
                                                     \frac{1}{2}
                                                     G_{ab}
                                                     \frac{\delta
                                                     \sigma_1}{\delta
                                                     h_a}\frac{\delta
                                                     \sigma_1}{\delta
                                                     h_b} +
                                                     \frac{1}{2} G_{ab}\frac{\delta^2
                                                     \sigma_1}{\delta
                                                     h_a \delta h_b} +
                                                     \frac{1}{2}g_a\frac{\delta
                                                     \sigma_1}{\delta
                                                     h_a}= 0$ \\ 
\hline 
$M^{-2}$ & $ G_{ab}
                                                   \frac{\delta
                                                   \sigma_0}{\delta
                                                   h_a}\frac{\delta
                                                   \sigma_3}{\delta
                                                   h_b}+G_{ab}
                                                   \frac{\delta
                                                   \sigma_1}{\delta
                                                   h_a}\frac{\delta
                                                   \sigma_2}{\delta
                                                   h_b} + 
                                                   \frac{1}{2}G_{ab}\frac{\delta^2
                                                   \sigma_2}{\delta
                                                   h_a \delta h_b}+
                                                   \frac{1}{2}g_a\frac{\delta
                                                   \sigma_2}{\delta
                                                   h_a} = 0$ \\ 
\hline
$M^{-3}$ & $G_{ab} \frac{\delta
                                       \sigma_0}{\delta
                                       h_a}\frac{\delta
                                       \sigma_4}{\delta h_a}+  G_{ab}
                                       \frac{\delta \sigma_1}{\delta
                                       h_a}\frac{\delta
                                       \sigma_3}{\delta
                                       h_b}+\frac{1}{2}G_{ab}
                                       \frac{\delta \sigma_2}{\delta
                                       h_a}\frac{\delta
                                       \sigma_2}{\delta h_b}  + 
                                       \frac{1}{2}G_{ab}\frac{\delta^2
                                       \sigma_3}{\delta h_a \delta
                                       h_b} +
                                       \frac{1}{2}g_a\frac{\delta
                                       \sigma_3}{\delta h_a}= 0$ \\ 
\hline 
\end{tabular}
 \caption{Equations for the $\sigma_j$ at consecutive orders of
   $M^{-1}$. One recognizes that the highest order only enters via the
 $\tau$-derivative.}
 \label{tab:C5_TableWKBOrders}
 \end{table}
}
Plugging (\ref{eq:BO_Ansatz}) into (\ref{eq:WDW-Equation}), we find
\begin{equation}
 \frac{\hbar^2}{M}\frac{1}{\chi}G_{ab}\frac{\delta \chi}{\delta
   h_a}\frac{\delta \psi}{\delta h_b} = H_{\rm m} \psi + \rho_\chi  \psi -
 \frac{\hbar^2}{2M}\left(G_{ab}\frac{\delta^2 \psi}{\delta h_a \delta h_b} +
   g_a \frac{\delta \psi}{\delta
     h_a}\right) \label{eq:GeneralEquationpsi}.
\end{equation}
With the choice $\rho_\chi=0$, we can now obtain from (\ref{eq:GeneralEquationpsi}) with the
definition of the time derivative in (\ref{eq:DefTimeDerivative})  and
the definition of $P$ in 
(\ref{eq:WKBChi}) an equation for the time evolution of $\psi$: 
\begin{equation}
 {\rm i}\hbar \frac{\delta \psi}{\delta \tau} = H_{\rm m} \psi - {\rm i}\hbar
 G_{ab}\frac{\delta P}{\delta h_a} \frac{\delta \psi}{\delta h_b} -
 \frac{\hbar^2}{2M}\left( G_{ab}\frac{\delta^2 \psi}{\delta h_a \delta h_b}
   + g_a \frac{\delta \psi}{\delta
     h_a}\right) \label{eq:FunctionalSchroedinger}. 
\end{equation}
One can proceed further by demanding that the gradient of $\psi$ is
proportional to the gradient of $\sigma_0$ as suggested in
\cite{KS91}, which is some sort of adiabatic approximation in
superspace; that is,
\begin{equation}
 \frac{\delta \psi}{\delta h_a} = \alpha[h] \frac{\delta
   \sigma_0}{\delta h_a} \label{eq:ParallelTime}, 
\end{equation}
for some functional $\alpha[h]$. Equation (\ref{eq:HamiltonJacobi}) then
yields $\frac{\delta \psi}{\delta \tau} = -2V\alpha$, and we can
express $\frac{\delta \alpha}{\delta \tau}$ in terms of
$\tau$-derivatives of $\psi$ and $V$. We further assume that there is
a total (not necessarily hermitian) Hamiltonian $H$ such that we can write 
\begin{equation}
{\rm i}\hbar \frac{\delta \psi}{\delta \tau} = H
\psi \label{eq:SchroedingerEquationAnsatz}. 
\end{equation}
Using this definition together with (\ref{eq:VanVleck}) and expressing the
second time-derivatives in (\ref{eq:FunctionalSchroedinger})
with the Hamiltonian by using again 
(\ref{eq:SchroedingerEquationAnsatz}), equation
(\ref{eq:FunctionalSchroedinger}) reduces to the following compact expression: 
\begin{equation}
 {\rm i}\hbar \frac{\delta \psi}{\delta \tau} = H \psi =  H_{\rm m} \psi -
 \frac{1}{4MV}\left(H^2 + {\rm i}\hbar \frac{\delta H}{\delta \tau}
 -{\rm i}\hbar K H\right)
 \psi \label{eq:ResultNonUnitary}, 
\end{equation} 
where 
\begin{equation}
K := \frac{1}{V}\frac{\delta V}{\delta \tau} -\frac{2{\rm i}M}{\hbar}
\sum_{j=2}^\infty \left(\frac{\hbar}{{\rm i}M} \right)^j \frac{\delta
  \sigma_j}{\delta \tau}. 
\end{equation}
This equation is the full functional
Schr\"odinger equation for $\psi$ {\em including all quantum gravitational
corrections} within the vacuum gravity ($\rho_\chi=0$)
approximation. Note that the Hamiltonian $H$ appears on both sides of
(\ref{eq:ResultNonUnitary}), so it is also an implicit equation
for the Hamiltonian itself. As we see, it correctly reproduces the
limit of quantum 
field theory in a fixed curved background spacetime at zeroth order in
$M^{-1}$. The equations at any other order in $M^{-1}$ can be found from 
(\ref{eq:ResultNonUnitary}) in a straightforward way by iteration.

We see that this equation is not only an equation for the dynamics of
$\psi$, but can also be used for determining the Hamiltonian $H$
itself. For simple cosmological models one could try to solve 
(\ref{eq:ResultNonUnitary}) for $H$ as a differential equation in
time. Within a semiclassical approximation, one could assume that $H$
is a function of $H_{\rm m}$ alone, which resembles a restriction to the
one-particle sector of the theory \cite{KS91}, and further
expand $H$ and $H_{\rm m}$ in $M^{-1}$. This approach is in fact in the
spirit of the traditional Born-Oppenheimer approximation
\cite{BornOppenheimer27}. For instance, the first order correction to
(\ref{eq:ResultNonUnitary}) is 
\begin{eqnarray}
 {\rm i}\hbar \frac{\delta \psi}{\delta \tau} &=& H_{\rm m} \psi -
 \frac{1}{4MV}\left(H_{\rm m}^2 + {\rm i}\hbar \frac{\delta H_{\rm m}}{\delta
                                       \tau} -
   \frac{{\rm i}\hbar}{V} \frac{\delta V}{\delta \tau} H_{\rm m}\right) \psi +
 \mathcal{O}(M^{-2}) \nonumber\\
&=& H_{\rm m} \psi -
    \frac{H_{\rm m}^2}{4MV} \psi-\frac{\hbar}{4M}\frac{\delta}{\delta\tau}
\left(\frac{{\rm i}H_{\rm m}}{V}\right) \psi+\mathcal{O}(M^{-2})
\label{eq:ResultNonUnitaryFirstOrder}, 
\end{eqnarray}
as presented in \cite{KS91}. It should be noted that
(\ref{eq:ResultNonUnitary}) is independent of any factor
ordering ambiguities, as the $g_a$-term has canceled due to 
(\ref{eq:VanVleck}). It is also interesting to note that the same
expression, equation (\ref{eq:ResultNonUnitary}), is obtained if we
assume that all $\sigma_i$, $V$, and $\psi$ only depend on $\tau$
such that we consider only the $\tau$ and $\tau$-$\tau$ components of
$g_a$ and $G_{ab}$ and drop all other components from the beginning in
equation (\ref{eq:WDW-Equation}). This is at first glance rather
surprising, and seems to be a direct consequence of 
(\ref{eq:VanVleck}), whose form in either case assures that all
additional terms are canceled. However, we see that 
(\ref{eq:HamiltonJacobi}) has a solution of the form $\tau=V$ and
$\sigma_0=-\int V^2{\rm d}^3x$ (recall that $G_{\tau \tau}=-{1}/{2V}$). In this
case, $\sigma_0$ and $V$ indeed only depend on $\tau$.

\subsection{Problems with unitarity}

\label{subsec:ProblemsUnitarity}
We recognize from (\ref{eq:ResultNonUnitary}) that the gauge choice
$\rho_\chi=0$ leads to a non-unitary time evolution for $\psi$ at
order $M^{-1}$ and higher. In
applications to cosmological models, the unitarity violating terms
have often been neglected as they are small compared to the unitary
terms \cite{Kiefer12,Bini13,BKK16b}. We can see two reasons for
the occurrence of non-unitarity directly in this equation. First, the Wheeler-DeWitt
equation is a Klein-Gordon type of an equation, whose unitary time
evolution (with repect to the usual quantum mechanical inner product) is spoiled
due to the second time derivatives, giving rise to the  term ${\rm i}\hbar
\frac{\delta H}{\delta \tau}$ in (\ref{eq:ResultNonUnitary}); see
equation (3.13) in \cite{honnef93}.
Secondly, the choice $\rho_\chi=0$ leads to the term ${\rm i}\hbar KH$.

The major problem at this stage is then the physical interpretation of
$\psi$ with a non-unitary time evolution: is $\psi$ really the
physical quantity described by the functional Schr\"odinger equation
plus quantum gravitational corrections? If yes, the gauge choice
$\rho_\chi=0$ has physical meaning above all other physical
requirements on our semiclassical limit. Then unitarity violation is
physical for $\psi$. Yet, the only important guiding principles we
have is to recover the correct classical limit (Hamilton-Jacobi
equation) and the functional Schr\"odinger equation in a fixed curved
background. But the condition $\rho_\chi=0$ leads to a set of additional
equations at each order in $M^{-1}$, whose interpretation is not clear.
In the next section we therefore show how one can, in
principle, render the theory unitary by modifying the gauge choice and
allowing $\rho_\chi\ne 0$.

\section{Unitary functional Schr\"odinger equation and non-vacuum
  gravity background ($\rho_\chi\ne0$)} 
\label{sec:UnitaryPreservation}

Following the standard Born-Oppenheimer approximation (see section~2 and
\cite{BK98,KTV17}), we shall in this section
demand unitary time evolution for the matter wave function
$\psi$, where unitarity is defined with respect to the standard
Schr\"odinger inner product. Since the full wave functional is given
as a solution to the Wheeler-DeWitt equation, this will modify both
$\psi$ and the gravitational wave function $\chi$ compared to the last
section. We shall see that this leads to different quantum
gravitational corrections at order $M^{-1}$. 

The main procedure is simple: we
adjust the matter wave functional such that we get, order by order, a unitary
evolution in WKB time. For simplicity, we will assume in this section that
$V$ and the $\sigma_i$ (and thus $\chi$) only depend on the WKB time $\tau$
defined in (\ref{eq:DefTimeDerivative}), and that
$\psi=\psi(\tau,\phi)$, and $H_{\rm m}=H_{\rm m}(\tau,\phi)$. This captures the
essential 
point of unitarity restoration; for the general case, one has to
include the terms discussed in \cite{BK98}. In this section, we set
$\hbar=1$ for simplicity. The Wheeler-DeWitt equation then reduces to
the 
following equation: 
\begin{equation}
 \left[ -\frac{1}{2M}\left(G_{\tau \tau}\frac{\delta^2}{\delta \tau^2}
     + g_\tau \frac{\delta}{\delta \tau}\right)+ MV(\tau) +
   H_{\rm m}(\tau,\phi)\right]\Psi = 0 \label{eq:WDW-Equation_onlytime}. 
\end{equation}
This form is obtained by an orthogonal decomposition of the
DeWitt metric as presented in \cite{Padmanabhan90}. 
For the total wave functional, we make an ansatz similar to (\ref{eq:BO_Ansatz}),
\begin{equation}
\Psi[\tau,\phi]=\chi[\tau]\psi[\tau,\phi].
\end{equation}
We assume $\Psi$ to be of the form
\begin{equation}
 \Psi = \exp\left({\rm i}M\sigma_0[\tau]\right) \psi_0[\tau,\phi] =:
   \chi_0 \psi_0 \label{eq:UnitaryStep1},
\end{equation}
and demand that $\rho_{\chi_0}=\mathcal{O}(M^0)$, where now, similar
to the last section, $$\rho_\chi := \frac{1}{\chi} \left[
  -\frac{1}{2M}\left(G_{\tau \tau}\frac{\delta^2}{\delta \tau^2} 
+ g_\tau \frac{\delta}{\delta \tau}\right)+ MV(\tau) \right]\chi.$$
This leads to the 
Hamilton-Jacobi equation (\ref{eq:HamiltonJacobi}) for
$\sigma_0$ at order $M$; it also allows us to keep the
definition (\ref{eq:DefTimeDerivative}) for $\tau$; hence, $G_{\tau \tau} =
-{1}/{2V}$ as before. With this new $\rho_{\chi_0}$, 
we find instead of (\ref{eq:ResultNonUnitary}),
\begin{equation}
 {\rm i} \frac{\delta \psi_0}{\delta \tau} \equiv H \psi_0=  H_{\rm m} \psi_0
 - {\rm i} \left(
   \frac{1}{2V} \frac{\delta V}{\delta \tau} -V g_\tau \right)\psi_0 -
 \frac{1}{4MV}\left(H^2 + {\rm i} \frac{\delta H}{\delta \tau} - 2{\rm
     i}Vg_\tau H
 \right) \psi_0 \label{eq:UnitaryStep2}.  
\end{equation}

At this stage, $\psi_0$ experiences a non-unitary time evolution as before.
In order to remedy this, we could split the Hamiltonian $H$
into hermitian and anti-hermitian parts, $H = H_{\rm H} + H_{\rm N}$,
and redefine the wave functional such that the new one evolves unitarily.
In case that the Hamiltonian commutes at different times, this can
 be achieved by
$\psi = \exp({\rm i} \int {\rm d} \tau H_{\rm N})\psi_0$, leading
to ${\rm i} {\delta
  \psi}/{\delta \tau} = H_{\rm H} \psi$. The wave functional $\psi$ can then be
interpreted as a physical wave functional. This would, however, only
be possible
if we knew $H$, which is in general not the case. We will thus
follow here an alternative route and show
how (\ref{eq:UnitaryStep2}) can be solved iteratively by demanding
unitarity for the matter wave functions {\em at each order}. 

Let us define two functions
$E(\tau)$ and  $\varepsilon(\tau)$ such that $H
\psi_0 = E \psi_0$ and $H_{\rm m} \psi_0 = \varepsilon \psi_0$;
$E$ is in general complex, but $\varepsilon$ is always
real.
The second equation means that we have to solve the matter
Schr\"odinger equation and 
take $\psi_0$ as an eigenfunction of $H_{\rm m}$. Let us further expand these
functions in the spirit of Born and Oppenheimer as 
\begin{equation}
E = E^{(0)} +
M^{-1}E^{(1)} + M^{-2}E^{(2)} + \dots, 
\end{equation}
 and similarly for $\varepsilon$. Inserting these expansions into 
(\ref{eq:UnitaryStep2}) yields for the first two orders: 
\begin{eqnarray}
 E^{(0)}&=&\varepsilon^{(0)} - {\rm i}\left( \frac{1}{2V} \frac{\delta
            V}{\delta \tau} -V g_\tau \right)  \nonumber \\ 
 E^{(1)}&=&\varepsilon^{(1)} - \frac{1}{4V}\left((E^{(0)})^2 + {\rm i}
            \frac{\delta E^{(0)}}{\delta \tau} - 2{\rm i}Vg_\tau E^{(0)}
            \right) \label{eq:UnitaryOrders}; 
\end{eqnarray}
in the last line, the expression for $E^{(0)}$ should be plugged in to
get an expression for $E^{(1)}$ in terms of $\varepsilon^{(0)}$ and
$\varepsilon^{(1)}$. Next we define 
\begin{equation}
 \psi_1 := \exp \left(- \int  \Im(E^{(0)}){\rm d} \tau\right)\psi_0=
 \exp\left(\int \left( \frac{1}{2V} \frac{\delta V}{\delta \tau} -V
     g_\tau \right) {\rm d} \tau \right) \psi_0 \label{eq:Unitarypsi1}, 
\end{equation}
so that $\psi_1$ obeys a unitary Schr\"odinger equation \emph{at order
  $M^0$}, that is, ${\rm i} \frac{\delta \psi_1}{\delta \tau} = H_{\rm
  m} \psi_1 +
\mathcal{O}(M^{-1})$. The total wave function thus reads
\begin{equation}
 \Psi = \chi_0 \psi_0 = \exp\left({\rm i}M\sigma_0-\int \left( \frac{1}{2V}
     \frac{\delta V}{\delta \tau} -V g_\tau \right) {\rm d} \tau \right)
 \psi_1 =: \chi_1 \psi_1. \label{eq:UnitaryStep3} 
\end{equation}
Note that if $g_\tau=0$, one can integrate the exponent to yield $\chi_1
= \frac{1}{\sqrt{V}}\exp ({\rm i}M\sigma_0)$, which is similar to the usual
WKB case (except for the difference mentioned in footnote~10). One can now
compute $\rho_{\chi_1}$ and finds that the zeroth order of
$\rho_{\chi_1}$ vanishes, $\rho_{\chi_1} =
\mathcal{O}(M^{-1})$. Hence, equation (\ref{eq:VanVleck}) (with only
$\tau$-derivatives) remains unchanged. This could have been anticipated by
noting that the unitary violating terms in 
(\ref{eq:ResultNonUnitary}) are of order $\mathcal{O}(M^{-1})$ and
higher. If we write, similar to our discussion above, $\chi_1 =
\exp({\rm i}M\sigma_0 + \sigma_1)$, we can read off (\ref{eq:VanVleck})
directly from the exponent of (\ref{eq:UnitaryStep3}):
$$\frac{\delta \sigma_1}{\delta \tau} = V g_\tau-\frac{1}{2V}
\frac{\delta V}{\delta \tau}.$$

In the next order, we define in an analogous way
\begin{eqnarray}
 \psi_2 &:=& \exp \left(-\frac{1}{M}\int \Im (E^{(1)}){\rm d} \tau
             \right)\psi_1 \nonumber \\ 
 &=& \exp \left(\frac{1}{M}\int \left(- 
     \varepsilon^{(0)}\left[\frac{1}{4V^2}\frac{\delta V}{\delta \tau} -
     \frac{1}{2}g_\tau \right] + \frac{1}{4V}\frac{\delta
     \varepsilon^{(0)}}{\delta \tau} -\frac{1}{2}g_\tau
     \varepsilon^{(0)}\right){\rm d} \tau \right)\psi_1 \nonumber \\ 
 &=& \exp \left(\frac{1}{4MV}\varepsilon^{(0)}(\tau)
     \right)\psi_1 \label{eq:Unitarypsi2} ,
\end{eqnarray} 
where in the last step we have performed a partial integration and omitted
a constant. We see that the $g_\tau$-term has dropped out in
this step. We hence get a new expression for the total wave function: 
\begin{equation}
 \Psi = \exp \left( {\rm i}M\sigma_0-\int \left( \frac{1}{2V} \frac{\delta
       V}{\delta \tau} -V g_\tau \right) {\rm d} \tau -
   \frac{1}{4MV}\varepsilon^{(0)}(\tau) \right) \psi_2 =: \chi_2 \psi_2. 
\end{equation}
If we define again $\chi_2 =\exp ({\rm i}M\sigma_0 + \sigma_1 - {\rm
  i}M^{-1}\sigma_2)$, we get for $\sigma_2$ the surprisingly simple
result: 
\begin{equation}
 {\rm i}\sigma_2 = \frac{1}{4V}\varepsilon^{(0)}(\tau) \label{eq:CorrectedSigma_2},
\end{equation}
so $\sigma_2$ is purely imaginary. This is different from the
situation of the last section.

One can easily see that at this order $M^{-1}$, $\rho_{\chi_2}$ does
not vanish. Requiring unitary time evolution for $\psi_2$
leads to the occurrence of a back reaction of the matter part onto the
gravitational sector, cf. \cite{BK98,KTV17}.
This is closer in spirit to the traditional
Born-Oppenheimer scheme than the approach discussed in the previous
section, as we will briefly discuss now. 

Let us consider equation (\ref{eq:WDW-Equation_onlytime}) with $G_{\tau
  \tau}=-{1}/{2V}$ and $\Psi = \chi[\tau] \psi[\tau,\phi]$. The
original approach of Born and Oppenheimer was to multiply this
equation with $\psi^*$ from the left and integrating over $\phi$. We
now require unitary time evolution for $\psi$. We define a real-valued function
$\bar{E}(\tau)$ by ${\rm i} \frac{\delta \psi}{\delta \tau} =\bar{E}
\psi$, $\bar{E}$ being real as $\psi$ is assumed to evolve
unitarily. This leads to the standard form
for the `nuclear' wave function: 
\begin{equation}
 \left[\frac{1}{4MV}\left(\frac{\delta}{\delta \tau} - {\rm i}
     \bar{E}\right)^2 -\frac{1}{2M} g_\tau \left(\frac{\delta}{\delta
       \tau}-{\rm i}\bar{E} \right) + \varepsilon + MV \right] \chi
 =0 \label{eq:UsualBO}. 
\end{equation}

At this stage, the large gauge group $e^{A[\tau]}$ has been reduced to the
unitary group, that is, $A$ is purely imaginary. We see that 
$\bar{E}$, which is the effective energy of the matter degrees of
freedom, assumes the role of the Mead-Berry connection in
this approach, which is different from the
usual approach in molecular physics.\footnote{Hence, employing the
  Born-Oppenheimer \emph{approximation}, i.e. dropping the Mead-Berry
  connection, will not help us here.} 
Note that, even though in
general the Mead-Berry connection can be gauged away by a unitary
transformation (by switching to the
diabatic picture, see 
e.g. \cite{Baer06}), this make no sense here, for we want to
determine $\bar{E}$ in this particular frame. 

As above, we can now make an ansatz of the form (also compare \cite{KTV17} for such a WKB-like ansatz within the traditional Born-Oppenheimer approach)
\begin{equation}
 \chi = \exp \left( {\rm i}M\eta_0 + \eta_1 + M^{-1}\eta_2 + \dots
 \right) \label{eq:UsualBOWKBAnsatz}, 
\end{equation}
where $\eta_{0,1} = \sigma_{0,1}$, $\eta_2 = -{\rm i} \sigma_2$, etc., and
all $\eta_i$ are taken to be real; the phase of $\psi$ is hence
influenced at the lowest order only. Expanding $\bar{E}$ in powers of $M^{-1}$,
plugging this ansatz into (\ref{eq:UsualBO}), and matching the
real and imaginary parts at each order, we find at order $M^{1}$ again
the Hamilton-Jacobi equation (\ref{eq:HamiltonJacobi}). At the
next order $M^0$, the real part of equation (\ref{eq:UsualBO}) yields $\bar{E}^{(0)} =
\varepsilon^{(0)}$, while the imaginary part yields an equation for the
Van Vleck determinant similar to (\ref{eq:VanVleck}), 
\begin{equation}
 \frac{\delta \eta_1}{\delta \tau} = g_\tau V -
 \frac{1}{2V}\frac{\delta V}{\delta \tau} \label{eq:UsualBOVanVleck}. 
\end{equation}
At the next order, we find for the imaginary part equation
(\ref{eq:CorrectedSigma_2}), while the real part yields  
\begin{equation}
\bar{E}^{(1)} = \varepsilon^{(1)}-\frac{1}{4V}(\varepsilon^{(0)})^2 +
\frac{1}{4V}\left(\frac{3}{4V^2}\left(\frac{\delta V}{\delta
      \tau}\right)^2 - V^2 g_\tau^2 - \frac{1}{2V}\frac{\delta^2
    V}{\delta \tau^2} + V \frac{\delta g_\tau}{\delta \tau} + g_\tau
  \frac{\delta V}{\delta \tau} \right) \label{eq:E1}, 
\end{equation}
which we would also get from (\ref{eq:UnitaryOrders}) if
we considered only the real part. Hence, as expected, the two methods
lead us to the same result if we require unitary time evolution for
$\psi$. At order $M^{-1}$, the energy $\varepsilon$, introduced above as
the eigenvalue corresponding to $H_{\rm m}$, is thus shifted by quantum
gravitational corrections; this shift is given by (recall that a
factor $M^{-1}$ must be added to the terms on the
right-hand side in (\ref{eq:E1}) to get the energy)
\begin{equation}
 \Delta\varepsilon = \frac{1}{4MV}\left(-\varepsilon^2 +
   \frac{3}{4V^2}\left(\frac{\delta V}{\delta \tau}\right)^2 - V^2
   g_\tau^2 - \frac{1}{2V}\frac{\delta^2 V}{\delta \tau^2} + V
   \frac{\delta g_\tau}{\delta \tau} + g_\tau \frac{\delta V}{\delta
     \tau}\right) + \mathcal{O}(M^{-2}) \label{eq:EnergyShift}. 
\end{equation}
The only relevant term is the contribution proportional to
$\varepsilon^2$, because the other terms are matter-independent. 
This relevant term is, in fact, the term that 
was used to calculate the
quantum gravitational correction to the power spectrum of the cosmic
microwave background (CMB) anisotropies \cite{Bini13,BKK16b,Kiefer12}.
In these papers, the first method was used and the unitarity-violating
terms were neglected by hand. This procedure can be justified by the
discussion presented here.

\section{Discussion}

Let us summarize the main results of our paper.
We have investigated semiclassical (Born-Oppenheimer type of)
approximation schemes for the 
Wheeler-DeWitt equation of canonical quantum gravity. The analogous situation in molecular physics was reviewed with an emphasis on the gauge freedom that arises within this framework. Although
the total entangled quantum state of electrons and nuclei is always
the same, this gauge freedom allows to shift terms between the
electronic and nucleonic parts. Requiring unitarity separately for the
electronic part, we get a definite expression for the back reaction onto the
nuclei; this is the usual Born-Oppenheimer approximation. 

The Born-Oppenheimer like ansatz for the total wavefunctional
entails a similar gauge freedom for the Wheeler-DeWitt equation. As in
molecular physics, a straightforward expansion in terms of the inverse
Planck-mass squared without back reaction of matter on
gravity spoils unitarity in the matter
sector.\footnote{One should note, however, that the analogy between
  these two cases is limited, since we are dealing with two
  fundamentally different equations -- the Schr\"odinger equation and
  the Wheeler-DeWitt equation, which is of Klein-Gordon type. The
  latter obeys a conservation law different from the former
  \cite{honnef93}.}
 We have discussed this scheme and extended it to all orders in the
expansion parameter. We then have modified the expansion scheme by
using the gauge freedom to guarantee unitary evolution for the matter
sector. This is closer in spirit to the standard Born-Oppenheimer
scheme and 
leads to back reaction terms for the gravitational part. This point
of view was taken in \cite{BFV96,BK98,KTV17}. A concrete normalization
of the matter states
is not needed -- the important thing is the unitary development. 
We note that the issue of avoiding unitarity violating terms also occurs when
performing a non-relativistic expansion for the Klein-Gordon equation
in external electromagnetic and gravitational fields \cite{CL95}. 

A good understanding of the semiclassical approximation to quantum
gravity is of fundamental importance for two main reasons. On the
theoretical side, it provides a bridge between full
quantum gravity and established physics. One can apply this scheme not
only to quantum general relativity with minimally coupled fields, but
also to scalar-tensor theories \cite{SW17}, Weyl gravity \cite{KN17},
and others. On the observational side, 
 first tests of quantum gravity will most
 likely occur from small correction terms that modify the usual limit
 of quantum field theory in curved spacetime \cite{BKK16a,BKK16b}.
 For this purpose, it is of great importance to
 develop and compare approximation schemes like the ones discussed here and
 investigate their empirical consequences.

\begin{acknowledgements}
We thank David Brizuela, Alexander Kamenshchik, Manuel Kr\"amer,
Branislav Nikoli\'c, Giovanni Venturi, and YiFan Wang for discussions and 
critical comments. 
\end{acknowledgements}


\begin{thebibliography}{99}

\bibitem{oup}
Kiefer, C.: Quantum Gravity. Oxford University Press, Oxford (2012)

\bibitem{grg}
Kiefer, C.: Gen. Rel. Grav. {\bf 41}, 877 (2009)

\bibitem{KS91}
Kiefer, C., Singh, T.P.: Phys. Rev. D {\bf 48}, 1067 (1991)

\bibitem{BFV96}
Bertoni, C., Finelli, F., Venturi, G.: 
Class. Quantum Grav. {\bf 13}, 2375 (1996)

\bibitem{BK98}
Barvinsky, A.O., Kiefer, C.: 
Nucl. Phys. B {\bf 526}, 509 (1998)

\bibitem{KTV17} Kamenshchik, A., Tronconi, A., Venturi, G.:
 Class. Quantum Grav. {\bf 35}, 015012 (2018)

\bibitem{Bini13}
Bini, D., Esposito, G., Kiefer, C., Kr\"amer, M., Pessina, F.:
Phys.Rev. D {\bf 87}, 104008 (2013)

\bibitem{BornOppenheimer27}
Born, M., Oppenheimer, R.: Ann. Phys. (Berlin) {\bf 389 (20)}, 457
(1927)

\bibitem{deco} Joos, E et al: Decoherence and the Appearance of a
  Classical World in Quantum Theory. Springer, Berlin (2003)

\bibitem{Bohm03}
Bohm, B., Mostafazadeh, A., Koizumi, H., Niu, Q., Zwanziger, J.: The
Geometric Phase in Quantum Systems. Springer, Heidelberg (2003) 

\bibitem{Baer06}
Baer, M.: Beyond Born-Oppenheimer -- Electronic Nonadiabatic Coupling
Terms and Conical Intersections. John Wiley \& Sons, Hoboken, New
Jersey (2006) 

\bibitem{Matyus} M\'atyus, E.: arXiv: 1801.05885 (2018)

\bibitem{BKK16a}
Brizuela, D., Kiefer, C., Kr\"amer, M.:
Phys. Rev. D {\bf 93}, 104035 (2016)

\bibitem{BKK16b}
Brizuela, D., Kiefer, C., Kr\"amer, M.:
Phys. Rev. D {\bf 94}, 123527 (2016)

\bibitem{Gerlach69}
Gerlach, U.H.: Phys. Rev. {\bf 177}, 1929 (1969)

\bibitem{Padmanabhan90}
Padmanabhan, T.: Pramana {\bf 35}, L199 (1990) 

\bibitem{Kiefer12}
Kiefer, C., Kr\"amer, M.: Phys. Rev. Lett. {\bf 108}, 021301 (2012)

\bibitem{honnef93}
Kiefer, C.: Lect. Notes Phys. {\bf 434}, 170 (1994)

\bibitem{CL95} L\"ammerzahl, C.: Phys. Lett.~A {\bf 203}, 12 (1995) 

\bibitem{SW17} Steinwachs, C. F., van der Wild, M. L.:
arXiv:1712.08543 (2017)

\bibitem{KN17} Kiefer, C., Nikoli\'c, B.:
Fundam. Theor. Phys. {\bf 187}, 127 (2017)



\end{thebibliography}
\end{document}